# The consistent behavior of negative Poisson's ratio with interlayer interactions

Yancong Wang[1], Linfeng Yu[1], Fa Zhang[1,2], Qiang Chen[1], Yuqi Zhan[1], Xiong Zheng[1,*], Huimin Wang[3,†], Zhenzhen Qin[4,‡], and Guangzhao Qin[1,§]

[1]*State Key Laboratory of Advanced Design and Manufacturing for Vehicle Body, College of Mechanical and Vehicle Engineering, Hunan University, Changsha 410082, P. R. China*
[2]*State Key Laboratory of Robotics and System, Harbin Institute of Technology, Harbin 150001, People's Republic of China*
[3]*Hunan Key Laboratory for Micro-Nano Energy Materials & Device and School of Physics and Optoelectronics, Xiangtan University, Xiangtan 411105, Hunan, China*
[4]*School of Physics and Microelectronics, Zhengzhou University, Zhengzhou 450001, China*

## Abstract

Negative Poisson's ratio (NPR) is of great interest due to the novel applications in lots of fields. Films are the most commonly used form in practical applications, which involves multiple layers. However, the effect of interlayer interactions on the NPR is still unclear. In this study, based on first principles calculations, we systematically investigate the effect of interlayer interactions on the NPR by comparably studying single-layer graphene, few-layer graphene, *h*-BN, and graphene-BN heterostructure. It is found that they almost have the same geometry-strain response. Consequently, the NPR in bilayer graphene, triple-layer graphene, and graphene-BN heterostructure are consistent with that in single-layer graphene and *h*-BN. The fundamental mechanism lies in that the response to strain of the orbital coupling are consistent under the effect of interlayer interactions. The deep understanding of the NPR with the effect of interlayer interactions as achieved in this study is beneficial for the future design and development of micro-/nanoscale electromechanical devices with novel functions based on nanostructures.

[*] Corresponding authors: X.Z. <xzheng@hnu.edu.cn>
[†] Corresponding authors: H.W. <wanghmin@xtu.edu.cn>
[‡] Corresponding authors: Z.Q. <qzz@zzu.edu.cn>
[§] Corresponding authors: G.Q. <gzqin@hnu.edu.cn>

# 1. Introduction

The Poisson's ratio, which varies from -1 to 0.5 on the basis of the classical elasticity theory[1], is one of the significant parameters to describe mechanic and physical properties. The existence of the negative Poisson's ratio (NPR) is also rational on theory. In recent years, NPR has been found in lots of materials, which are well known as auxetic materials. The NPR, which may enables many novel applications, attracts great interest because of the typically enhanced toughness, shear resistance, sound and vibration absorption[2]. In literature, there are extensive studies of NPR on bulk auxetic structures[3,4], metals[5,6], *etc*. Besides, several models have been proposed for the explanation[7–10]. Recently, the discovery of NPR in metal nanowires and carbon nanotubes has been reported.[11,12] In the study of two dimension (2D) materials, like the representative graphene, the NPR was discovered in 2D materials with specific engineering, such as introducing vacancy defects[13], creating periodic pores[14], cutting into nanoribbons[15], *etc*. In addition, the intrinsic in-plane NPR has been found in 2D materials when applying strain along a special direction without any external modification to the structure, shape or composition. For instance, the NPR has been recently discovered by prediction in 2D honeycomb structures of graphene, silicene, *h*-BN, *h*-GaN, *h*-SiC, and *h*-BAs[7]. Moreover, there are also some studies focusing on the out-of-plane NPR, such as TiN[16], phosphorene[17,18], arsenic[19,20], GeS[21], SnSe[22].

However, the studies in literature focus on the NPR in single-layer 2D materials, while limited studies have been done on the in-plane NPR of multi-layer 2D materials. In addition, The Poisson's ratio of bulk graphite is positive, which is quite different from the NPR of single-layer graphene, despite that the structure of bulk graphite can be viewed as a stack of many layers of single-layer graphene. As a result, the effects of interlayer interaction on the in-plane NPR are still not clear. In fact, there are lots of further researches to be conducted in this area. For instance, graphene film is usually used in reality instead of graphene, which involves multiple layers. Here, the bilayer graphene is a model for the study of layer effect. Moreover, the heterojunction structure with stacking of different materials is also an interesting topic. Besides, most of the previous explanations of NPR materials are based on the analysis of the evolution of geometric parameters, and only few studies have explored the mechanism at the electronic level[9,19]. Thus, it is necessary to study the effect of interactions between layers on the NPR of few-layers 2D materials and achieve a more fundamental understanding.

In this study, we systematically investigate the response of strain and key geometric parameters for bilayer graphene, triple-layer graphene, and graphene-BN heterostructure (gra-BN) with strain applied. It is found that the in-plane NPR is consistent among these structures while NPR in bilayer graphene and gra-BN is weakened in different degree. The mechanism is uncovered by analyzing the response of orbital coupling to strain. The results deepen the understanding of the NPR, which would shed light on future design of micro-/nanoscale electromechanical devices.

## 2. Methods

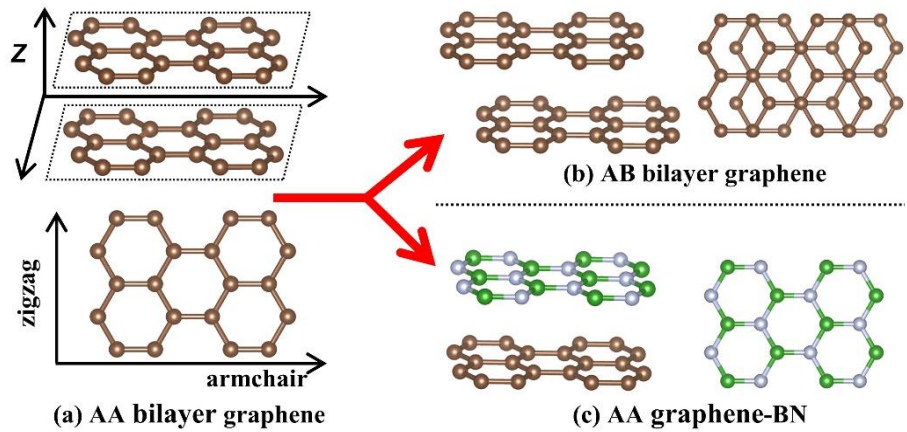

Figure 1. These schematics of the structures of bilayer graphene in (a) AA, (b) AB stacking and three-layer graphene in (c)ABA stacking.

All the calculations are performed using the Vienna *ab initio* simulation package (VASP)[23], which is based on density functional theory (DFT). The Perdew−Burke−Ernzerh[24] (PBE) of generalized gradient approximation (GGA) is chosen as the exchange-correlation functional. Long-range *van der Walls* interactions were taken into account using the optB88 vdW functional.[25,26] We use the classical single-layer graphene and two kinds of representative bilayer graphene with the same crystal orientation but different stacking (AA and AB stacking, as shown in Fig. 1). AA stacking corresponds to the stacking pattern of two single-layers paired with each other, while AB stacking corresponds to the stacking pattern moving one of the layers in the opposite direction for one-third of the [1,1] crystallographic vector[27]. The kinetic energy cutoff of wave functions is set as 1000 eV for all the calculations. The Monkhorst–Pack[28] k-mesh of 19 × 11 × 1 is used to sample the Brillouin zone (BZ), and the energy convergence threshold is set as $10^{-6}$ eV. Uniaxial strains along the typical zigzag and armchair directions are applied. The strain is defined as $(I - I_0)/I_0$, where $I$ is the lattice constant with stretching and $I_0$ is the original lattice constant without stretching. The stress is scaled by replacing the thickness including the vacuum space with the effective layer thickness. Specifically, for flat materials, the effective layer thickness is the sum of the actual distance between the two layers and the van der Waals diameter of the C atoms. [29–33]. In all cases, the geometric parameters are fully optimized with the Herman-Feynman force on all atoms less than $10^{-4}$ V/Å. The stability of the structures is verified by calculating the phonon dispersion (see Fig. S3 in the Supplementary Materials for more details). The optimized interlayer distances in AA, AB bilayer graphene and gra-BN are 3.55, 3.37 and 3.47Å, respectively. While without considering the vdW interlayer interaction, the optimized interlayer distance of the AA and AB bilayer graphene are 4.6 and 4.4 Å, respectively. This is much larger than the vdW diameter and can be viewed as a structural optimization of two separate single-layer graphene.

# 3. Results

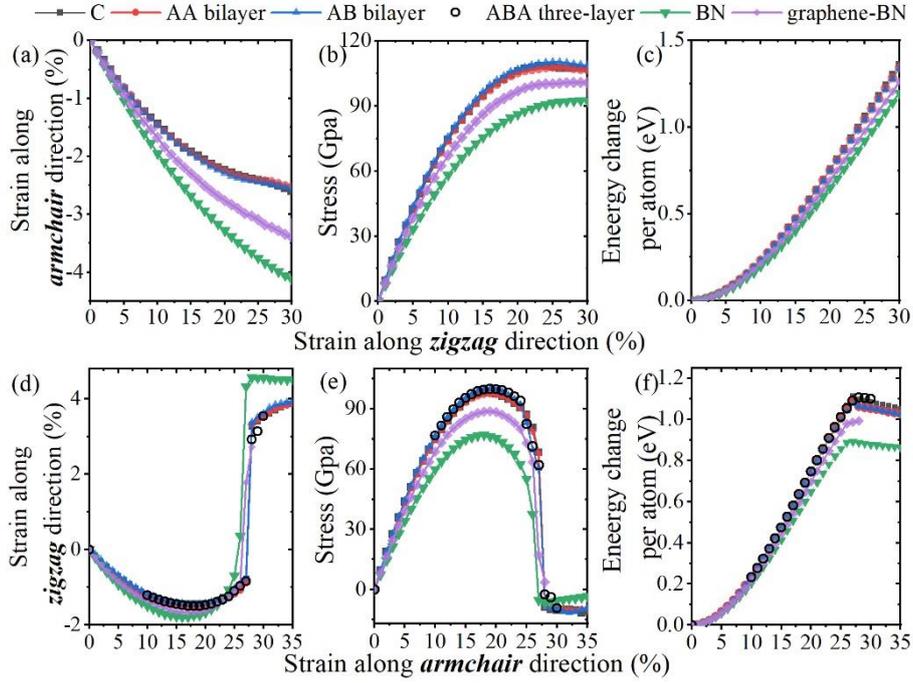

Figure 2. The anisotropic response of the (a, d) driven strain, (b, e) stress, and (c, f) energy change per atom with strain applied along the (a, b, c) zigzag and (d, e, f) armchair directions, respectively.

With strain applied, the mechanical response of single-layer graphene and bilayer graphene with AA and AB stacking are studied. Previous studies show that graphene can sustain a large strain (≥25%) and has a large breaking strength[34,35]. The breaking strength is found to be 42 N/m of graphene[34]. Considering that the effective thickness of graphene is about 0.35 nm, the fracture strength of 42 N/m graphene is 125.4 GPa, which is greater than the maximum stresses achieved along the two stretch directions in Fig. 2, 108 and 98 GPa. Generally speaking, from an experimental point of view, experimental results can be influenced by defects, temperature effects, or other elements that may react with the graphene layer. Because of the above reality, the theoretical value of the prediction is often higher than the experimental value. However, this does not affect the prediction of NPR of graphene materials and the exploration of its potential mechanism.

Fig. 2 shows the response of strain, stress and energy per atom for the single-layer graphene, AA/AB bilayer, ABA triple-layer graphene, BN and graphene-BN heterostructure. Generally, three physical parameters response to the strain along the zigzag direction are continuous.

However, with a significantly large strain applied, a mutant (28%) is found for the response to the strain along the armchair direction, which means the structure is failed when the strain along armchair direction is larger than 28%. It can be found that the response to the strain along zigzag direction in this study is a common phenomenon, and the lattice constant along armchair direction decreases with the increasing strain along the zigzag direction. While Fig. 2(d) shows that the lattice constant decreases when the strain along the armchair direction is between 0% and around 15%, the lattice constant increases abnormally when strain is larger than 15%, which indicates the appearance of NPR phenomena. The stress along the stretch direction increases except for the situations where NPR exist. The highest stress is found at the condition where the NPR starts to appear, and then the stress decreases with the increase of stretch. As for the energy per atom, it keeps growing regardless of the existence of negative Poisson effect. Because the strain in the stretching direction is positive, and so as to the stress, and there is no stress in the other direction. Then the positive work is being done and the energy is being put into the system. Moreover, the response to strain of the AA/AB bilayer graphene and single-layer graphene are almost the same, which demonstrates the consistency in the responses of strain, stress and energy of the three materials. Besides, the curve of the single-layer graphene is found to be slightly higher than bilayer graphene as shown in in Fig. 2(d), which is similar to the results reported in previous work on the study of multilayer graphene via molecular dynamics simulations.[36]

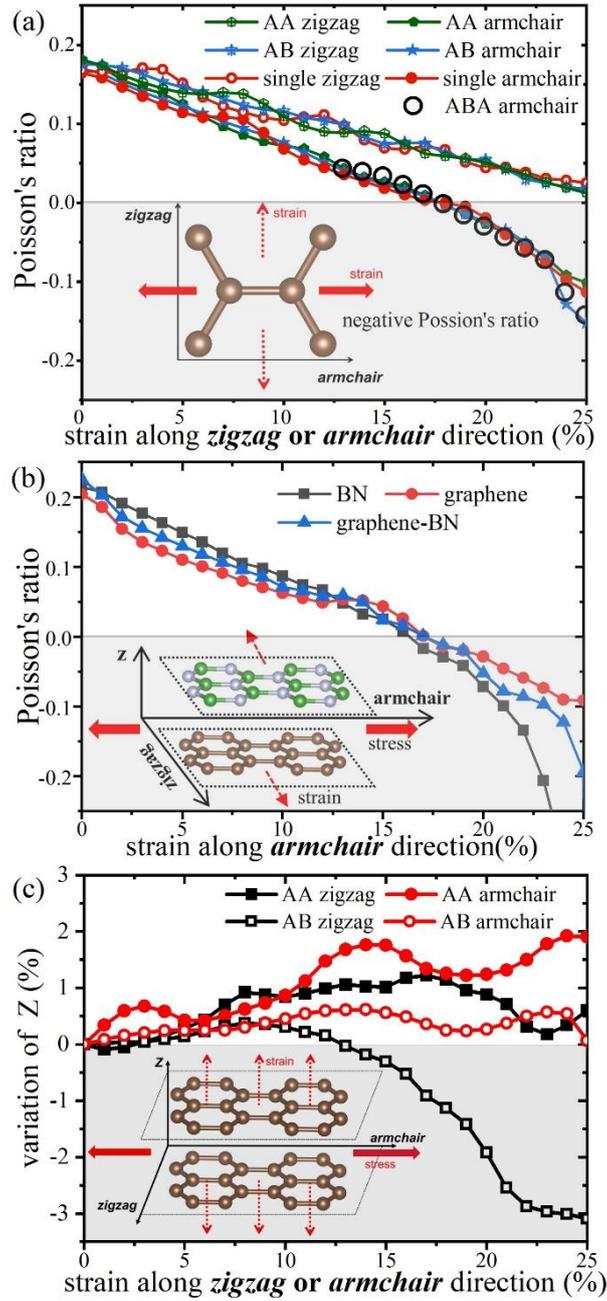

Figure 3. (a) The calculated Poisson's ratio of single-layer graphene, bilayer graphene in AA, AB and triple-layer graphene in ABA stacking. The inset shows that the NPR emerges with the expansion along the zigzag direction when the stretch strain is applied along the armchair direction. The NPR area are marked in gray. (b) The calculated Poisson's ratio of single-layer BN, graphene and graphene-BN heterostructure in AA stacking when strain is applied along armchair direction. (c) The variation of Z direction distance when strain is applied along zigzag or armchair direction. The inset shows that the distance between two layers increase when strain is applied, which reveals the *out-of-plane* NPR phenomena.

To get a more precise and intuitive view of the NPR, the Poisson's ratio is calculated and the results are shown in Fig. 3. It is well known that the Poisson's ratio is defined as[8]

$$\nu = -\frac{\partial \varepsilon_x}{\partial \varepsilon_y}, \qquad (1)$$

where $\nu$ is the Poisson's ratio, $\varepsilon_x$ is the strain along x direction and $\varepsilon_y$ is the strain along y direction. The x and y directions are perpendicular to each other, like the zigzag and armchair directions in this work. Fig. 3(a) shows that the Poisson's ratio of bilayer and single-layer graphene has almost the same value and trend. Therefore, the interlayer interactions in the structures of AA and AB bilayer graphene have little effect on the in-plane Poisson's ratio. Further, the Poisson's ratio of materials decreases faster when stretching along the armchair direction than zigzag direction. And the line of the descent is approximately a straight-like line, indicating that the slope changes very little, which is an interesting point. Such behavior is consistent with the variation of the lattice constant with strain applied. The Poisson's ratio becomes negative when the stretch is higher than around 15% along armchair direction, while the Poisson's ratio keeps positive during 0-25% stretch along zigzag direction. It's worth noting that the Poisson's ratio is almost negative when the stretch is 25% along the zigzag direction. Thus, it can be expected that NPR may exist when more strain is applied along the zigzag direction.[9]

With the results of single-layer and bilayer graphene, it is interesting to further investigate what happens in three-layer graphene. The black pentagons in Fig.3(a) show that the A-B-A stacking triple graphene basically possesses the same NPR as bilayer graphene. This is possibly because the third and first layers are far apart from each other. Thus, the interaction on in-plane NPR is weak and the NPR is consistent. Note that there is a sharp drop of the NPR in the ABA and AB stacking conditions as the stretching reaches 24%, which is probably because the structure has become less stable when the stretching reaches 25%.

It's worth mentioning that we also studied the NPR of heterostructures in Fig. 3(b). In terms of the large variation trend, the NPR of gra-BN is consistent with that of multilayer graphene. However, there are still some interesting differences. It is shown that the graphene-BN heterostructure significantly enhances the NPR of single-layer graphene after the NPR occurs in the stretching process along the armchair direction. In this process, the NPR effect of heterojunctions is still smaller than that of single-layer BN. However, before the NPR occurs, the situation is just the opposite: BN has the smallest NPR, single-layer graphene has the strongest NPR, and the heterostructure is in the middle position.

Moreover, the variation of the distance between layers ($Z$) is calculated when strain is applied. The results in Fig. 3(c) indicate interesting NPR phenomena along the *out-of-plane* direction, which will be useful for vertical vibration isolation applications of graphene-layered devices. Further study to explore mechanisms on this *out-of-plane* NPR phenomena needs to be conducted in future.

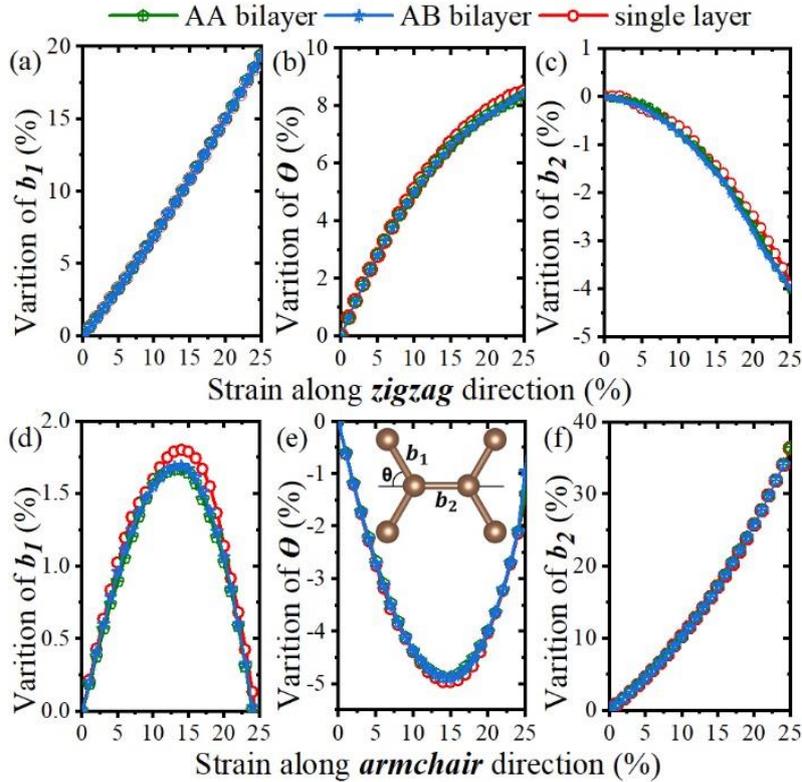

Figure 4. The evolution of the key geometry parameters ($b_1$, $b_2$ and $\theta$) with the strain applied along (a, b, c) zigzag and (d, e, f) armchair directions, respectively. The $b_1$, $b_2$ and $\theta$ are labeled on the inset of (e) for clarification.

To get insight into the variation process of inner geometric structure, we calculate the length of bond ($b_1$, $b_2$) and angle ($\theta$). As shown in Fig. 4, there is nearly no difference for single-layer graphene and AA/AB bilayer graphene. This indicates that the three materials have almost the same inner geometric variation during stretching, which gives rise to the same results to the strain in Fig. 2 and the NPR in Fig. 3. With further consideration, the interaction force between layers have nearly no influence on the in-plane variation. However, it is noticeable that single-layer graphene has slightly larger geometric structure change than that of bilayer graphene when the strain is applied in the armchair direction, especially with respect to $b_1$ and $\theta$ in Fig. 4(d, e). In addition, Fig. 4 shows the reason why the NPR phenomena appear and the different response to strain applied along different vertical directions at the geometric level. Fig. 4(a, b, c) shows that it's is monotonic that $b_1$ and $\theta$ increase, and $b_2$ decreases, resulting in the positive Poisson's ratio. While Fig. 4(d, e) shows that $b_1$ and $\theta$ increase first and then decrease, which is abnormal variation response to the lattice stretching. These abnormal variations result into the NPR. It is understandable that the increasing of $b_1$ and $\theta$ has opposite effect on the NPR. The increasing of $\theta$ causes the length of the zigzag increasing while the increasing of $b_1$ leads to the decreasing length of zigzag. Because the strain along the stretch direction is increasing statically, we can take only the lattice constant of another direction ($I_{zigzag}$ and $I_{armchair}$) into consideration. $I_{zigzag}$ and $I_{armchair}$ correspond to the lattice constants in two different stretch directions, respectively, which can be written as

$$I_{zigzag} = 2 \times b_1 \sin\theta , \qquad (2)$$

$$I_{armchair} = 2 \times b_2 + 2 \times b_1 \cos\theta, \qquad (3)$$

where $b_1$ and $b_2$ are the length of two bonds and $\theta$ is the value of bond angle as illustrated in Fig. 4, $I_{zigzag}$ and $I_{armchair}$ are the length of the lattice constant along the zigzag and armchair directions, respectively. Then, the actual $I_{zigzag}$ and $I_{armchair}$ have the same trend and change as the corresponding lattice constant strain in Fig. 2(a, d). Therefore, although the increasing $b_2$ strengths the NPR, the abnormal increasing of $\theta$ is the main factor for the occurrence of NPR.

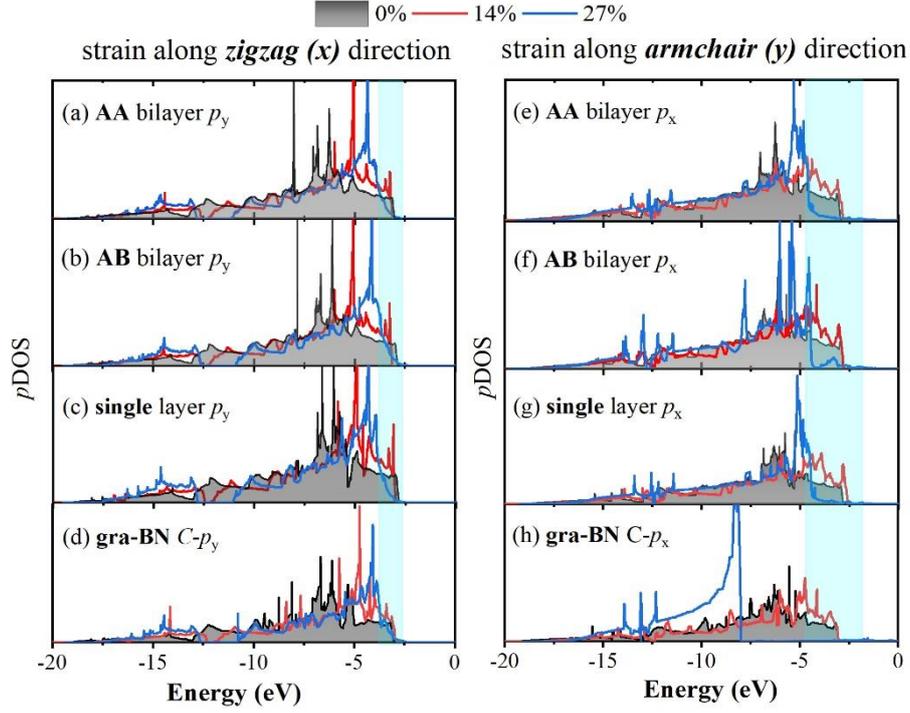

Figure 5. (a, b, c, d) The evolution of $p_y$ when strain is applied along the zigzag direction (x). (d, e, f, h) The evolution of $p_x$ when the strain is applied along the armchair direction (y). (a, e), (b, f), (c, g), (d, h) illustrate the situation of AA bilayer, AB bilayer, single-layer graphene and gra-BN, respectively. The cyan border highlights the key evolution around the valence band maximum.

To understand the internal mechanism of the NPR, we investigated the projected density of states ($p$DOS) to study the electronic function of three representative strength conditions (0%, 14%, and 27% corresponding to the cases with no strain, before NPR, and after NPR, respectively). With comparison of hybridized C-$p_x$/$p_y$ orbitals and the solo C-$p_z$ orbital closing to the valance band maximum (VBM) of the three different materials with two direction stretch (Fig. S1 in the Supplemental Materials), we find the recognized difference between the response to $p_x$ orbital (along zigzag direction) of stretch along the armchair direction and the response to $p_y$ orbital (along armchair direction) of stretch along the zigzag direction, casing the different response along the armchair and direction stretching. As shown in Fig. 5(e, f, g, h), four material's $p_x$-DOS closing to the VBM increase slightly first and then decrease significantly, which means the interaction force along the zigzag direction (x) decreases largely, causing $\theta$ increases abnormally in Fig. 4(e). This is the essential variation causing the NPR. While in Fig. 5(a, b, c, d), four materials' $p_y$-DOS closing to the VBM have the small variation and almost monotonic decline, which may cause the monotonic variation of geometric parameter in Fig. 4(a, b, c) and the response to positive passion's ratio.

Furthermore, we find that the $p_x$-DOS have almost the same trend between AA/AB bilayer graphene and single-layer graphene because most electrons are constrained to move in a 2D plane (see Fig. S2 in the Supplemental Materials for more information).[37] However, compared with single and bilayer graphene, it's worth to pointing out that whether the stretching is in the direction of armchair or zigzag, the $p$DOS of single-layer graphene changes the most, which is probably because the interlayer interaction in two-layer structure weakens the coupling in the in-plane direction slightly ($p_x$ or $p_y$). Especially, the degree of decline for the single-layer graphene has the largest reduction between 14% and 27% stretching along the armchair direction in Fig. 5(a, b and c), causing the obviously larger geometry changes ($b_1$ and $\theta$ in Fig. 4 (d, e)) than AA and AB bilayer graphene. In addition, Fig. 5(h) illustrates that the C- $p_x$ of gra-BN heterostructure has a huge decrease after the occurrence of NPR, which is much larger than that of single and bilayer graphene, and this is also the reason why the degree of NPR of gra-BN is larger than that of single and bilayer graphene.

## 4. Discussion and conclusions

In summary, by studying the strain, stress, energy per atoms and geometric responses to axial stretching with single-layer graphene, few-layer graphene, $h$-BN, and graphene-BN heterostructure, we found that the responses and NPR among them have almost the same behavior, indicating the weak effect of interlayer interactions on the in-plane NPR. This suggests that the difference between single-layer graphene and bulk graphite is a novel phenomenon of NPR in graphene caused by dimensional changes, which cannot be simply explained as the effect of interlayer interactions. In addition to the consistent behavior of the NPR, it is found that the variation of $\theta$ and $b_1$ in the single-layer graphene is larger than that in the bilayer graphene. Moreover, by studying the $p$DOS, it is indicated that $p_x$ slightly increases first and then decreases significantly during stretching along the armchair direction (y), which means that the interaction along the zigzag direction (x) decreases and then causes $\theta$ to increase abnormally, leading to in-plane NPR. In contrast, $p_y$ have slightly monopoly change during stretching along zigzag direction (x), and there is no NPR phenomenon. Moreover, the $p_x$ of single-layer graphene decreases slightly more than that of bilayer graphene and the $p_x$ of gra-BN heterostructure decreases largely more than that of single and bilayer graphene, which also leads to the greatest variation of geometric parameters ($\theta$ and $b_1$) in single-layer graphene and the greatest degree of NPR in gra-BN. It may be because that the interlayer interactions weaken the coupling of the in-plane $p_x$ orbital. Thus, the consistency of NPR in few-layer and single-layer graphene, and further BN and graphene-BN heterostructure is explained. The interlayer interactions may affect the *in-plane* coupling of *p*-orbitals slightly, leading to differences in the in-plane geometric change. Our study provides a deep understanding on the effect of interlayer interaction and reveal the internal mechanism of NPR in bilayer and single-layer graphene at the level of electron interaction. It is expected to shed light on future design and development of micro-/nanoscale electromechanical devices with novel functions based on nanostructures.

## Data Availability

The data that support the findings of this study are available from the corresponding authors upon reasonable request.

## Acknowledgements

The numerical calculations in this paper have been done on the supercomputing system of the National Supercomputing Center in Changsha. X.Z. is supported by the Fundamental Research Funds for the Central Universities (Grant No. 531118010490) and the National Natural Science Foundation of China (Grant No. 52006059). H.W. is supported by the National Natural Science Foundation of China (Grant No. 51906097). Z.Q. is supported by the National Natural Science Foundation of China (Grant No. 11847158, 11904324) and the China Postdoctoral Science Foundation (2018M642776). G.Q. is supported by the Fundamental Research Funds for the Central Universities (Grant No. 531118010471), the National Natural Science Foundation of China (Grant No. 52006057), and the Changsha Municipal Natural Science Foundation (Grant No. kq2014034). The authors declare that there are no competing interests.

## Author Contribution

G.Q. supervised the project. Y.W. and L.Y. performed all the calculations and analysis. All the authors contributed to interpreting the results. The paper was writing by Y.W. with contributions from all the authors.

*Supplementary Information*

**The consistent behavior of negative Poisson's ratio with interlayer interactions**

Yancong Wang[1], linfeng Yu[1], Fa Zhang[1,2], Qiang Chen[1], Yuqi Zhan[1], Xiong Zheng[1,*], Huimin Wang[3,*], Zhenzhen Qin[4,*], and Guangzhao Qin[1,*]

[1]*State Key Laboratory of Advanced Design and Manufacturing for Vehicle Body, College of Mechanical and Vehicle Engineering, Hunan University, Changsha 410082, P. R. China*
[2]*State Key Laboratory of Robotics and System, Harbin Institute of Technology, Harbin 150001, People's Republic of China*
[3]*Hunan Key Laboratory for Micro-Nano Energy Materials & Device and School of Physics and Optoelectronics, Xiangtan University, Xiangtan 411105, Hunan, China*
[4]*School of Physics and Microelectronics, Zhengzhou University, Zhengzhou 450001, China*

# 1. On the fundamental mechanism

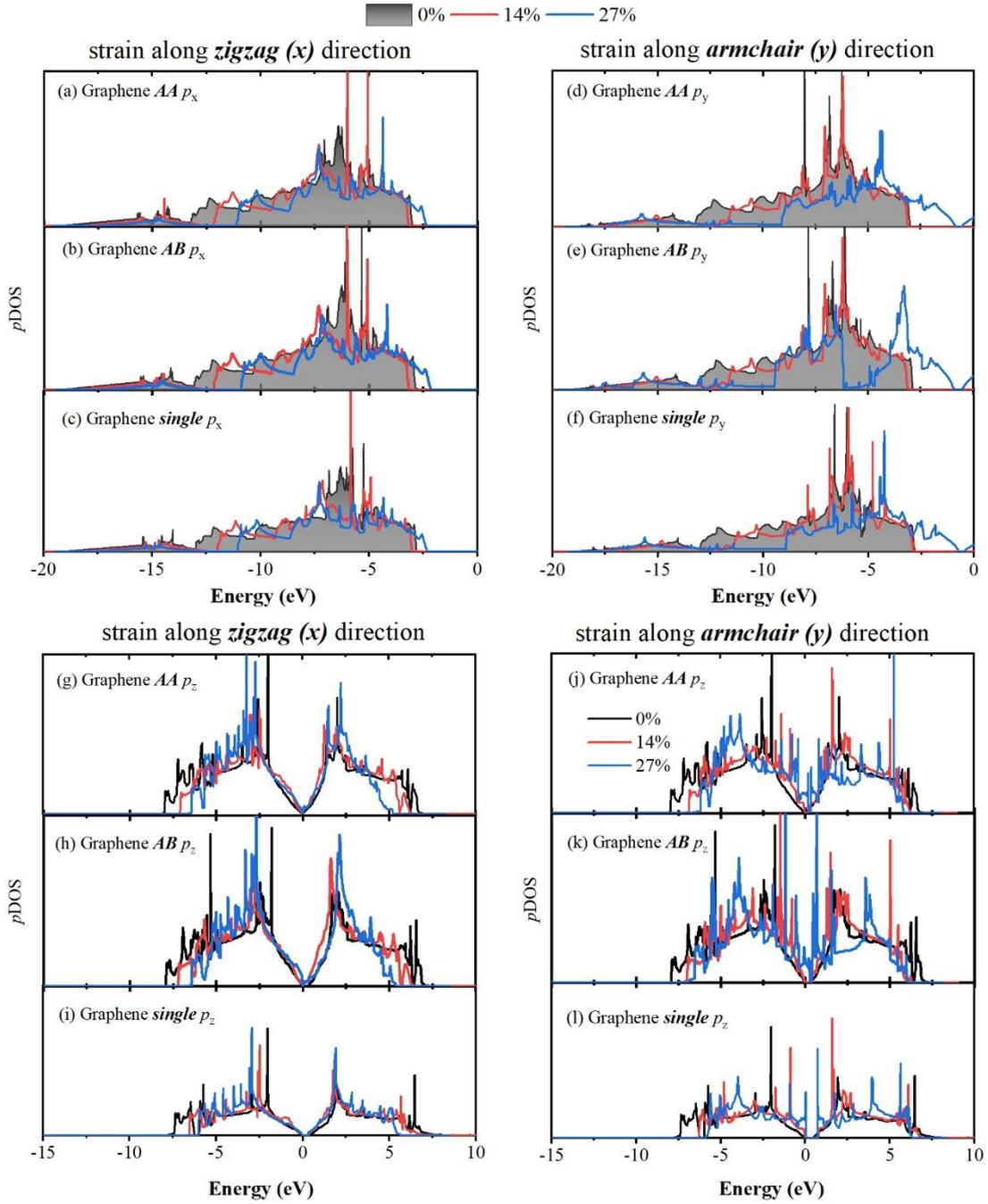

Figure S1: (a, b, c) The evolution of $p_x$ when strain is applied along the zigzag direction (x). (d, e, f) The evolution of $p_y$ when the strain is applied along the armchair direction (y). (g, h, i) The evolution of $p_z$ when the strain is applied along the zigzag direction (x). (j, k, l) The evolution of $p_z$ when the strain is applied along the armchair direction (y). (a, d, g, j), (b, e, h, k), and (c, f, i, l) illustrate the situation of AA bilayer, AB bilayer and single-lawyer graphene, respectively.

To get insight into the fundamental mechanism, we further study the evolution of orbital projected density of states ($p$DOS). It is well known that the C-C $\sigma$ bonds come from the hybridized C-$p_x$/$p_y$ orbitals, and the solo C-$p_z$ orbital forms the $\pi$ bonds and electronic Dirac cone.[1] Thus, we study $p_x$, $p_y$ and $p_z$ for bilayer and single-layer graphene when strain is applied. As shown in Figure S1, it has almost same value and trend in some hybridized orbitals. When strain is applied along zigzag direction (x), $p_x$ orbital decreased slightly then increased near the valance band maximum (VBM), which is almost same trend as $p_y$ when strain is applied along the armchair direction (y). That's probably because the direction of orbitals is same as the direction of uniform strain applied. Moreover, $p_z$ barely change near VBM during stretching. Thus, the different responses in orbital perpendicular to the direction of the strain applied cause different responses in NPR when strain is applied along zigzag and armchair direction.

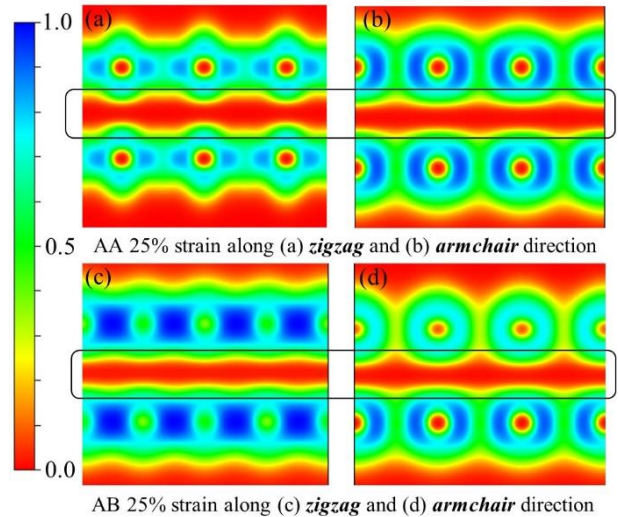

Figure S2: The evolution of the electronic localization functions (ELF) of (a, b) AA and (c, d) AB bilayer graphene when 25% strain is applied along (a, c) zigzag and (b, d) armchair direction, respectively. The areas framed by the black edges represent the interlayer position of AA and AB bilayer graphene.

As shown in Figure S2, there is low density of electrons in the interlayer position of AA and AB bilayer graphene probably because the electrons are restricted in in-plane area.[2] Thus, the interlayer interaction is weak and then have little effect on in-plane NPR, casing the consistency of bilayer and single-layer graphene. Furthermore, low density of electrons doesn't mean no electrons in the interlayer position of bilayer graphene. And these electrons unrestricted in 2D plane will slightly decrease the in-plane geometry variation on bilayer graphene.

## 2. Verification of the structural stability

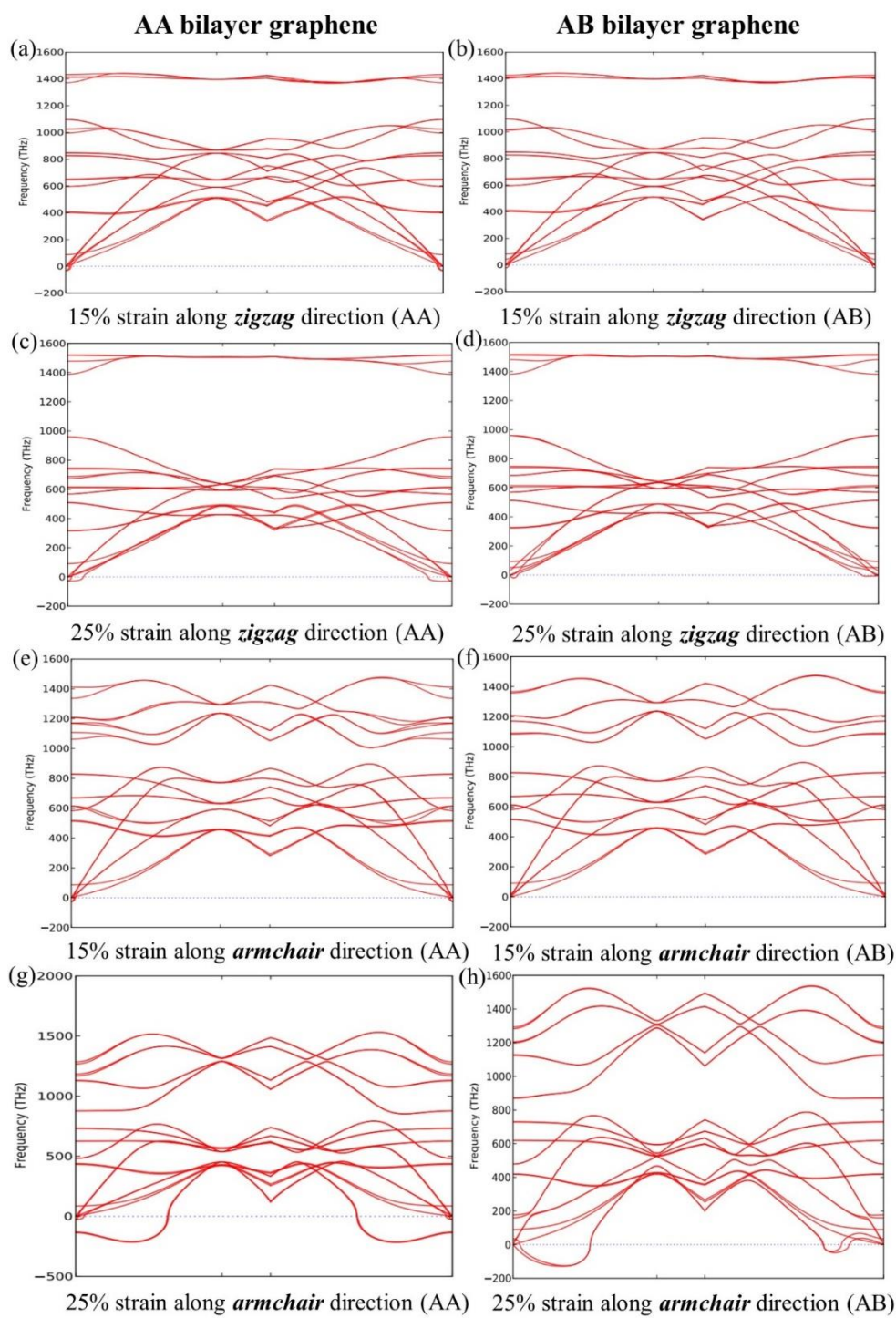

Figure S3: Phonon dispersions of (a, c, e, g) AA and (b, d, f, h) AB bilayer graphene when strain ((a, b, e, f) under 15% strain, (c, d, g, h) under 25% strain) is applied along (a, b, c, d) zigzag or (e, f, g, h) armchair direction, respectively.

We use phonon dispersions to verify the structural stability for AA and AB bilayer and single-layer graphene. As for single-layer graphene, it is the strongest material ever measured and it can sustain a large strain ( ≥25%)[3,4]. However, the mode has become imaginary in single-layer graphene when 25% strain is applied along armchair direction[1], as same as Fig. S3(g, h) in bilayer graphene under the same strain condition. Fig. S3 shows that AA and AB bilayer is stable when strain is applied along zigzag direction while the mode has become imaginary when strain near 25% is applied along armchair direction, probably because bilayer graphene structure break down near 30% strain is applied along armchair direction.